\documentclass[apjl]{aastex61}
\usepackage{epsfig,amssymb,amsmath,rotating,lineno,graphicx,cleveref,color}
\usepackage{color}
\newcommand\asat{{\it AstroSat}}
\newcommand\xte{{\it RXTE}}
\newcommand\swift{{\it Swift}}
\newcommand\inte{{\it INTEGRAL}}
\makeatletter
\newcommand\footnoteref[1]{\protected@xdef\@thefnmark{\ref{#1}}\@footnotemark}
\makeatother
\shorttitle{X-ray states during giant radio flare in Cyg X-3}
\shortauthors{Pahari et al.}
\begin{document}
\title{Extensive broadband X-ray monitoring during the formation of a giant radio jet base in Cyg X-3 with AstroSat}
\correspondingauthor{Mayukh Pahari}
\email{mayukh@iucaa.in}

\author{Mayukh Pahari}
\affiliation{Inter-University Center for Astronomy and Astrophysics, Ganeshkhind, Pune 411007, India}
\affiliation{Department of Astronomy and Astrophysics, Tata Institute of Fundamental Research,
Colaba, Mumbai 400005, India}

\author{J S Yadav}
\affiliation{Department of Astronomy and Astrophysics, Tata Institute of Fundamental Research,
Colaba, Mumbai 400005, India}

\author{Jai Verdhan Chauhan}
\affiliation{Department of Astronomy and Astrophysics, Tata Institute of Fundamental Research,
Colaba, Mumbai 400005, India}

\author{Divya Rawat}
\affiliation{Department of Physics, IIT Kanpur, Kanpur, Uttar Pradesh 208016, India}

\author{Ranjeev Misra}
\affiliation{Inter-University Center for Astronomy and Astrophysics, Ganeshkhind, Pune 411007, India}

\author{P C Agrawal}
\affiliation{UM-DAE Centre for Excellence in Basic Sciences, University of Mumbai, Kalina, Mumbai, Maharashtra 400098, India}

\author{Sunil Chandra}
\affiliation{Department of Astronomy and Astrophysics, Tata Institute of Fundamental Research,
Colaba, Mumbai 400005, India}
\affiliation{School of Physical and Chemical Sciences, North-West University, Potchefstroom 2520, South Africa}

\author{Kalyani Bagri}
\affiliation{School of Studies in Physics and Astrophysics, Pt. Ravishankar Shukla University, Raipur, Chhattisgarh 492010, India}

\author{Pankaj Jain}
\affiliation{Department of Physics, IIT Kanpur, Kanpur, Uttar Pradesh 208016, India}

\author{R K Manchanda}
\affiliation{University of Mumbai, Kalina, Mumbai 400098, India}

\author{Varsha Chitnis}
\affiliation{Department of High Energy Physics, Tata Institute of Fundamental Research,
Colaba, Mumbai 400005, India}

\author{Sudip Bhattacharyya}
\affiliation{Department of Astronomy and Astrophysics, Tata Institute of Fundamental Research,
Colaba, Mumbai 400005, India}

\label{firstpage}

\begin{abstract}

We present X-ray spectral and timing behavior of Cyg X-3 as observed by \asat{} during the onset of a giant radio flare on 01-02 April 2017. Within a time-scale of few hours, the source shows a transition from the hypersoft state (HPS) to a more luminous state (we termed as the very high state) which coincides with the time of the steep rise in radio flux density by an order of magnitude. Modeling the SXT and LAXPC spectra jointly in 0.5-70.0 keV, we found that the first few hours of the observation is dominated by the HPS with no significant counts above 17 keV. Later, an additional flat powerlaw component suddenly appeared in the spectra which extends to very high energies with the powerlaw photon index of 1.49$^{+0.04}_{-0.03}$. Such a flat powerlaw component has never been reported from Cyg X-3. Interestingly the fitted powerlaw model in 25-70 keV, when extrapolated to the radio frequency, predicts the radio flux density consistent with the trend measured from RATAN-600 telescope at 11.2 GHz. This provides a direct evidence of the synchrotron origin of flat X-ray powerlaw component and the most extensive monitoring of the broadband X-ray behavior at the moment of decoupling the giant radio jet base from the compact object in Cyg X-3. Using SXT and LAXPC observations, we determine the giant flare ejection time as MJD 57845.34 $\pm$ 0.08 when 11.2 GHz radio flux density increases from $\sim$100 to $\sim$478 mJy.

\end{abstract}


\section{Introduction}

Establishing the true nature of the connection between the accretion disk and radio jets in X-ray binaries at different mass accretion rate has been considered one of the challenging problems in astrophysics and several attempts have been made so far in this direction \citep{fe04,fe09,di14}. Such attempts successfully establish the long-term Radio/X-ray correlations from weeks to years time-scale in both transient and persistent/semi-persistent sources like GX 339-4 \citep{co13}, Cyg X-1 \citep{st01}, GRS 1915+105 \citep{mu01,fe04a}, Cyg X-3 \citep{zd16}. However, there are a handful of attempts where inner disk activity was closely monitored before, during and after the superluminal radio ejection from X-ray transients. \citet{mi12} made a remarkable attempt during the 2009 outburst of the Galactic micro-quasar H 1743-322. They monitored the formation and ejection of superluminal radio jets using VLBA at 8.4 GHz between 28 May and 6 June 2009 almost on a daily basis in simultaneous with the \xte{}/PCA observations to monitor accretion flow properties. They found that Radio emission was highly quenched before the major radio flare and the ejection event occurred during the state transition from hard-intermediate to soft state which is marked by the disappearance of type-C quasi-periodic oscillations (QPOs) from the power density spectra \citep{ca05}.      
Until the present, the changes in X-ray spectro-temporal properties during the formation of a superluminal jet has been monitored with the frequency of days to weeks except GRS 1915+105 which is only source that has radio and sensitive X-ray data on day scale \citep{ya06}. However, depending upon the size of the radio jet base, it is possible that the decoupling may occurs in first few minutes to hours and may leave a signature in X-ray spectro-timing properties. Unfortunately, there exist no studies where the change in X-ray properties of the inner accretion flow are traced with the accuracy of minutes time-scale. One key reason for the lack of such studies is that superluminal radio ejections are rare, transient and their time of occurrences are completely unpredictable within days to week time-scale. Although a few studies with microquasars have shown that major radio flares occur during the transition from the hard-intermediate to the soft spectral state \citep{fe09}, the occurrence of such transition is highly uncertain within week time-scales. Therefore, it is difficult to determine the time of radio jet formation more accurately than few days.

Cyg X-3 is a Galactic, high mass, wind-fed X-ray binary \citep{pa72} and on many occasions it shows unique giant radio flares with the peak Radio flux density up to 20 Jy \citep{le75,mo88,sc95,mi01}. Additionally radio and X-ray variability have been observed from this source, and their correlation has been studied \citep{zd16}. Based on hardness and X-ray \& radio flux, five X-ray spectral states have been identified \citep{ko10}: hypersoft state (HPS), flaring soft X-ray state (FSXR), flaring intermediate state (FIM), flaring hard X-ray state (FHXR) and quiescent state. QPOs of the order of mHz are detected from Cyg X-3 during the hard state \citep{pa17,ko11,ma93}.
The latest giant radio flare which reached nearly 20 Jy on 04-05 April 2017 was first predicted by \citet{tr17a}. Monitoring Cyg X-3 using RATAN-600 radio telescope in three bands: 4.6 GHz, 8.2 GHz, and 11.2 GHz on a daily basis and comparing the \swift{}/BAT hard X-ray flux, they showed that before the giant flare, the source entered into the HPS where both hard X-ray and radio flux are highly quenched \citep{ko17}. On 1-2 April 2017, the radio flux started increasing rapidly, and on 04-05 April it reached the peak value of $\sim$20 Jy in all three radio bands. During this period, 15-50 keV \swift{}/BAT flux increases by a factor of $\sim$2-3. Similar transition from the ultrasoft to the hard X-ray state is also observed previously with \inte{} \citep{be07}. Interestingly, \asat{} observed the source continuously on 01-02 April when the X-ray state transition and major radio flare formation was taking place.

\begin{figure*}
\begin{center}
\includegraphics[scale=0.33,angle=-90]{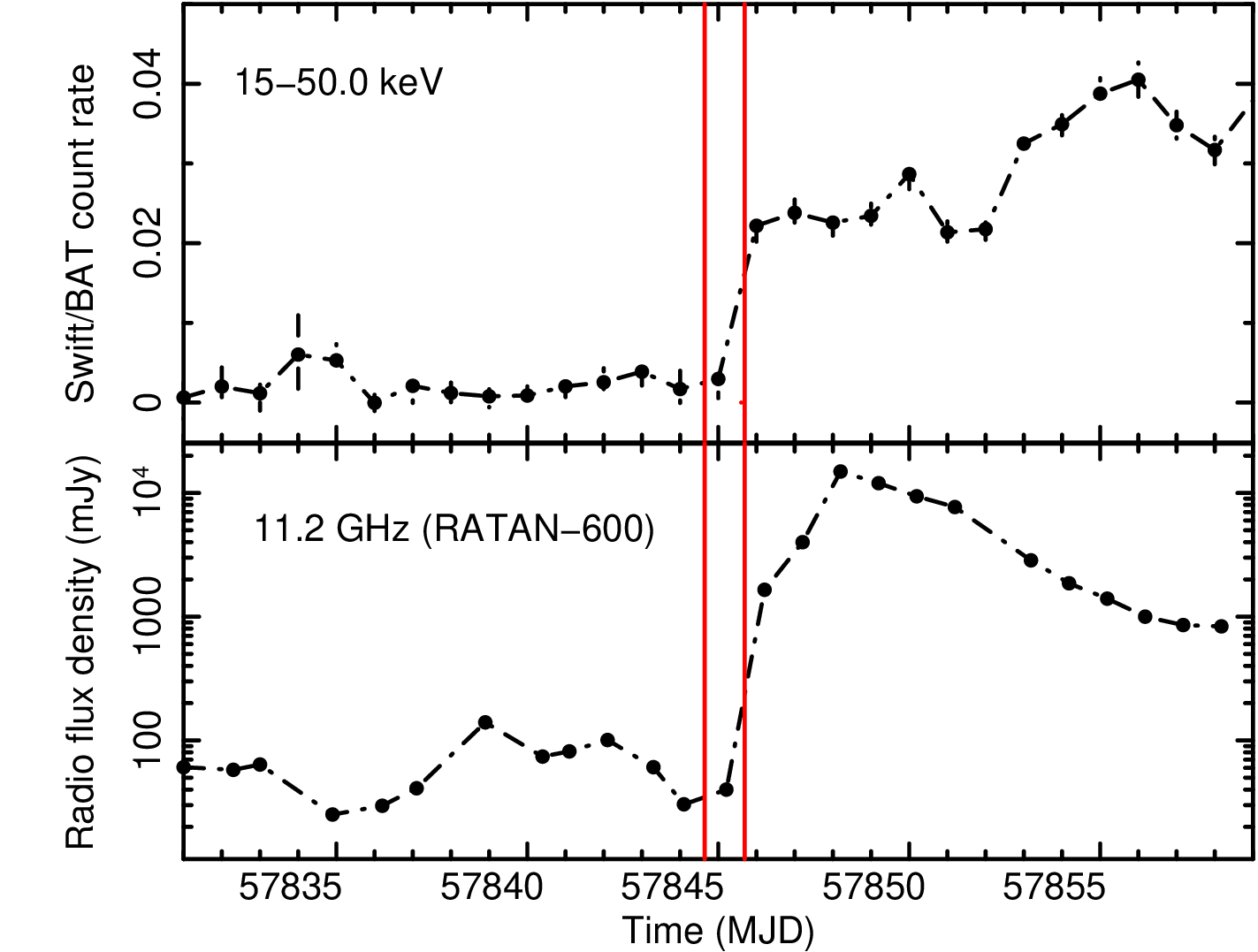}
\includegraphics[scale=0.33,angle=-90]{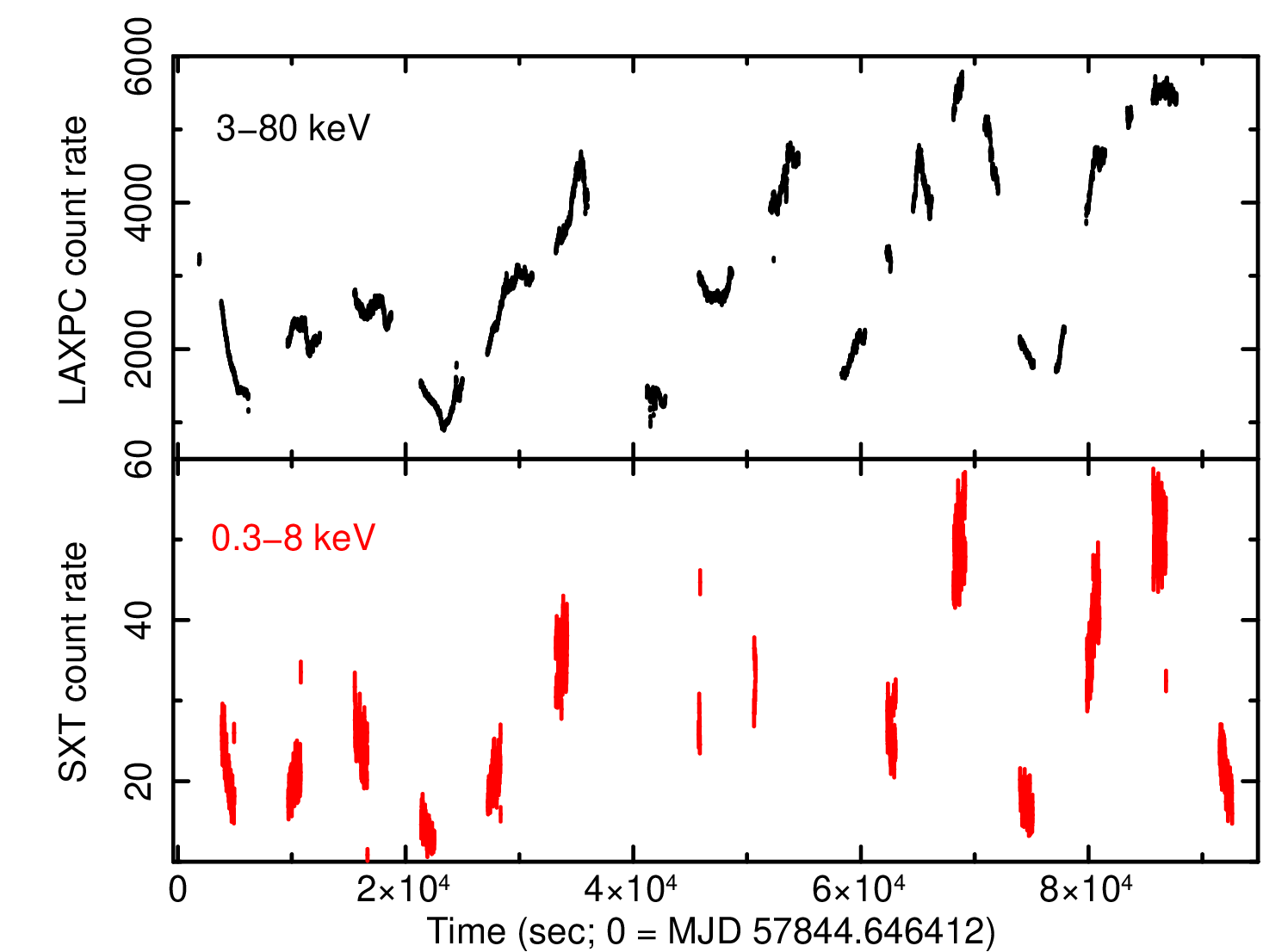}
\includegraphics[scale=0.33,angle=-90]{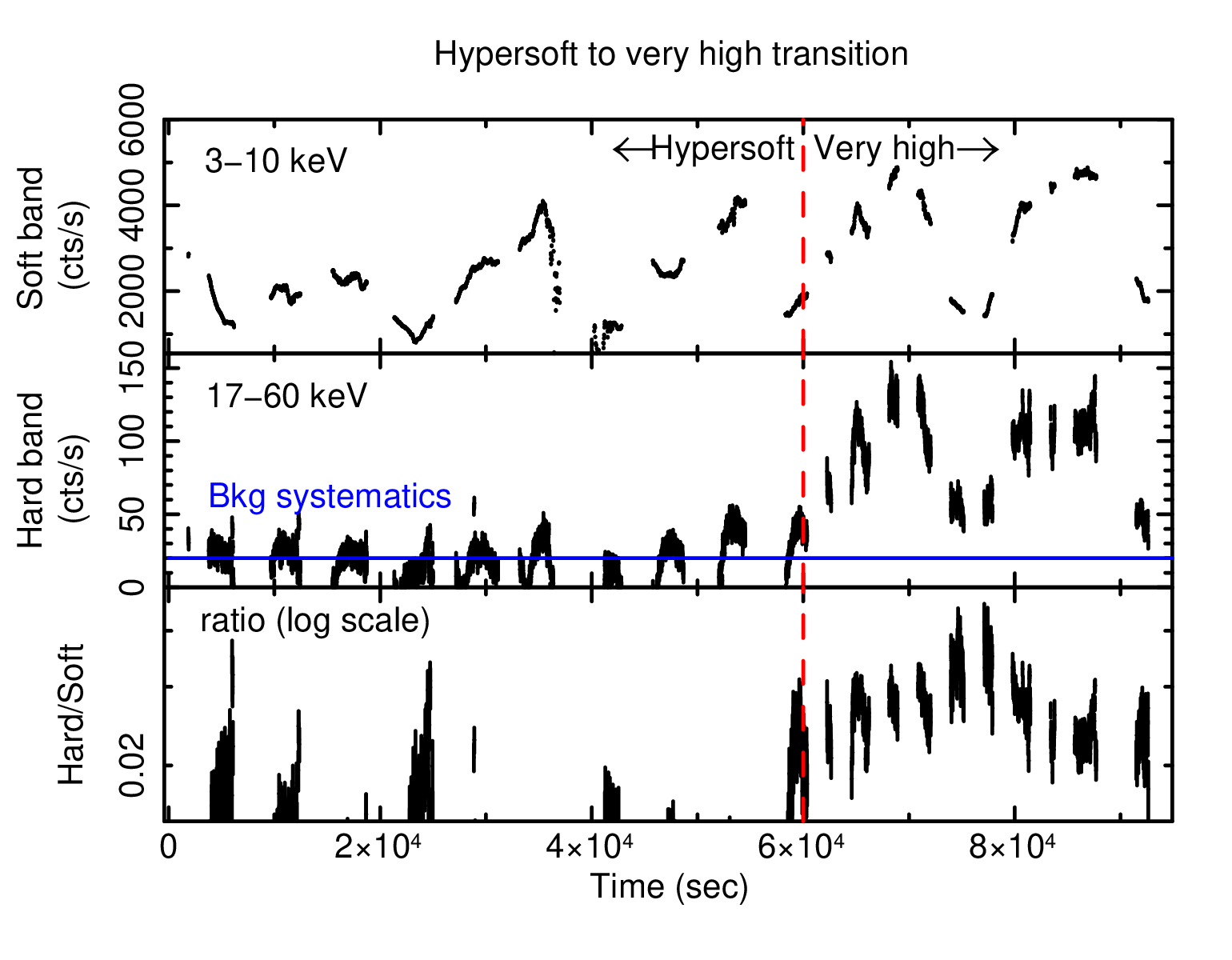}
\includegraphics[scale=0.33,angle=-90]{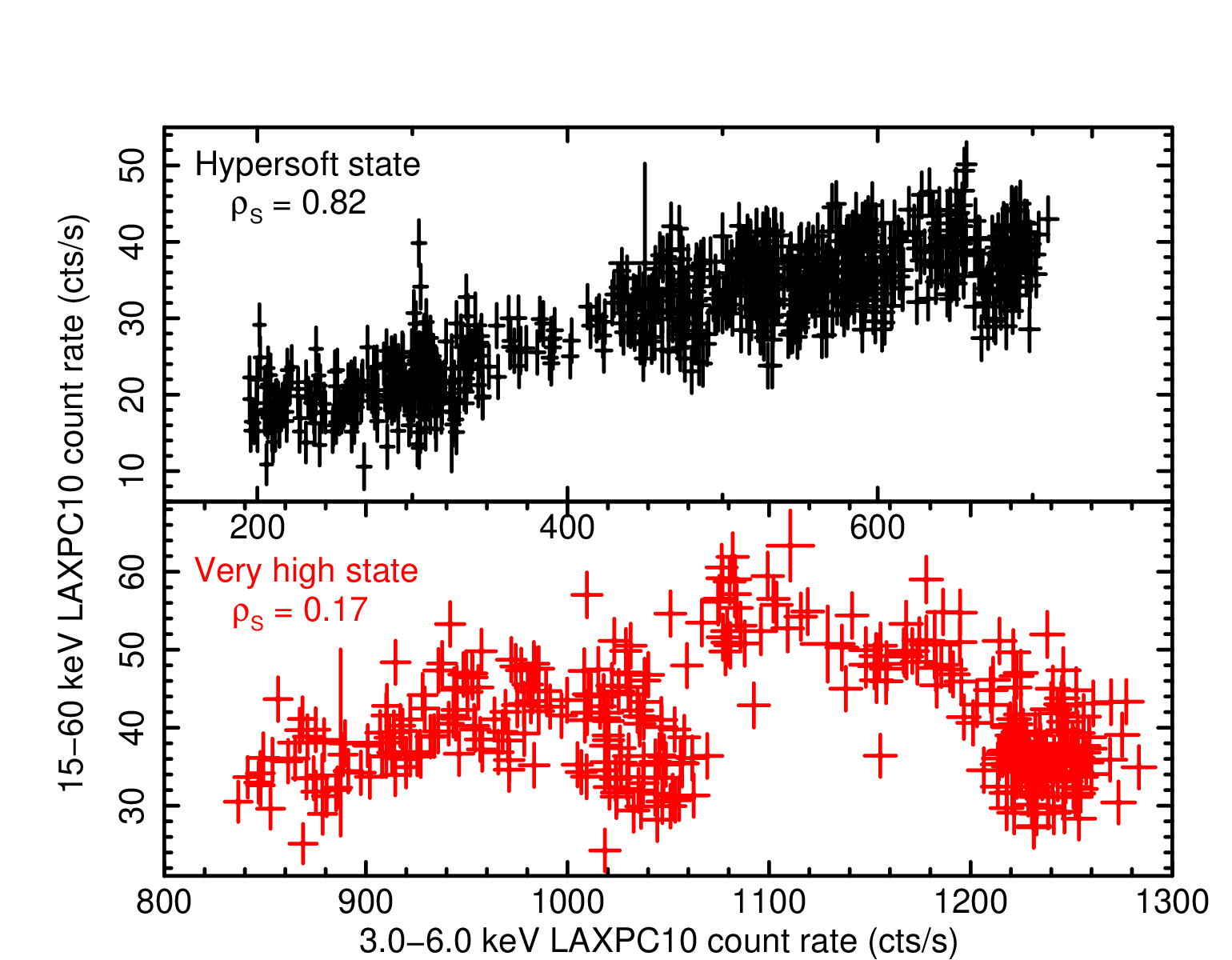}
\caption{Top left panel shows Swift/BAT lightcurve of Cyg X-3 in the 15-50 keV (top) and simultaneous RATAN-600 radio lightcurve (bottom) at 11.2 GHz (adopted from \citep{tr17a,tr17b}). The time interval between two vertical red lines is the observation period of \asat{}. During this period, top right panel shows 3-80 keV background-subtracted LAXPC lightcurve combining all three units (top) and 0.3-8 keV SXT lightcurve (bottom). For clarity, 10 sec bin time is used. Bottom left panel shows the background-subtracted LAXPC lightcurve in 3-10 keV (top), 17-60 keV (middle) and their ratio. The dotted vertical line separate two spectral states: hypersoft and very high. The blue horizontal line marked the 5\% systematics errors on the estimated background. Bottom right panel show flux-flux plot between 3-6 keV and 15-60 keV during the hypersoft (top) and the very high (bottom) state where $\rho_s$ denotes the Spearman rank correlation co-efficient.}  
\label{light}
\end{center}
\end{figure*}

In this work, we analyze the simultaneous observations of Cyg X-3 using Soft X-ray Telescope (SXT) and Large Area X-ray Proportional Counter (LAXPC) instruments on-board \asat{}. Based on lightcurve, hardness, and broadband energy spectra, we found that during the first 10-12 hours, the source was in the HPS. After $\sim$50 ks, an unusual flat powerlaw component appeared in the spectrum of Cyg X-3 and we termed it as the very high state (VHS). We show that the flat powerlaw component is consistent with the synchrotron emission from the radio jet monitored simultaneously with RATAN telescope. This observation provides an opportunity to closely monitor the X-ray properties during the formation of giant radio flare in Cyg X-3.

\begin{table*}
 \centering
 \caption{Best-fit parameters using SXT and LAXPC joint spectral fitting during the hypersoft and the very high state}
\begin{center}
\scalebox{0.85}{%
\begin{tabular}{ccccccccc}
\hline 
model & parameters & Hypersoft & Very High  \\
 & & state & state  \\
\hline 
{\tt tbabs} & N$_{H,ISM}$ (10$^{22}$ cm$^{-2}$) & 3.17$^{+0.04}_{-0.05}$ & 3.63$^{+0.04}_{-0.03}$  \\
{\tt pcfabs} & N$_{H,loc}$ (10$^{22}$ cm$^{-2}$) & 13.5$^{+0.5}_{-0.6}$ & 14.1$^{+0.4}_{-0.5}$ \\
&                  f$_{sc}$ & 0.65$^{+0.02}_{-0.01}$ & 0.71$^{+0.02}_{-0.02}$ \\
{\tt diskbb} & kT$_{in}$ (keV) & 1.39$^{+0.03}_{-0.02}$ & 1.11$^{+0.03}_{-0.02}$ \\
              & $\sqrt{N_{dbb}}$ & 16.4$^{+3.1}_{-2.6}$ & 31.8$^{+2.1}_{-4.4}$ \\
{\tt bremss}    & kT$_{plasma}$ (keV) & 4.58$^{+0.42}_{-0.31}$ & 3.87$^{+0.19}_{-0.16}$ \\
                        & N$_{bremss}$  & 0.59$^{+0.21}_{-0.18}$ & 3.62$^{+0.62}_{-0.53}$ \\
{\tt powerlaw} & $\Gamma$ & -- & 1.43$^{+0.06}_{-0.03}$ \\
               & N$_{powerlaw}$ & -- & 0.04$^{+0.01}_{-0.01}$ \\
{\tt gauss}   & E$_{gauss}$ (keV) & 6.94$^{+0.04}_{-0.03}$ & 7.08$^{+0.04}_{-0.12}$ \\
              & W$_{eq}$ (eV) & 96$^{+11}_{-8}$ & 59$^{+5}_{-3}$ \\
{\tt gabs}   & E$_{gabs}$ (keV) & 2.96$^{+0.01}_{-0.01}$ & 2.95$^{+0.01}_{-0.02}$ \\
                    & $\sigma_{gabs}$ (keV) & 0.07$^{+0.01}_{-0.01}$ & 0.06$^{+0.02}_{-0.01}$ \\
                   & $Depth_{gabs}$ (keV) & 0.04$^{+0.01}_{-0.01}$ & 0.03$^{+0.01}_{-0.01}$ \\
F$_{0.3-3}$   & (10$^{-10}$ ergs s$^{-1}$ cm$^{-2}$) & 8.68$^{+0.22}_{-0.43}$ & 13.6$^{+0.64}_{-0.62}$ \\
F$_{3-6}$   & (10$^{-10}$ ergs s$^{-1}$ cm$^{-2}$) & 53.4$^{+1.29}_{-1.55}$ & 55.8$^{+1.69}_{-1.43}$ \\
F$_{6-10}$   & (10$^{-10}$ ergs s$^{-1}$ cm$^{-2}$) & 15.8$^{+1.13}_{-0.56}$ & 28.7$^{+1.14}_{-1.15}$ \\
F$_{10-20}$   & (10$^{-10}$ ergs s$^{-1}$ cm$^{-2}$) & 2.32$^{+0.13}_{-0.09}$ & 7.91$^{+1.09}_{-1.17}$ \\
F$_{20-80}$   & (10$^{-10}$ ergs s$^{-1}$ cm$^{-2}$) & 0.35$^{+0.12}_{-0.24}$ & 9.43$^{+3.17}_{-2.32}$ \\
              & $\chi^2$/dof & 981/708 (1.39) & 1221/854 (1.43) \\
\hline
\end{tabular}}
\tablecomments{N$_{H,ISM}$ and N$_{H,loc}$ are the line-of-sight, inter-stellar medium and local absorption column density respectively, f$_{sc}$ is the scattering fraction, kT$_{in}$ and N$_{dbb}$ are the inner disk temperature and {\tt diskbb} normalization respectively. $\Gamma$ and N$_{powerlaw}$ are photon powerlaw index and the {\tt powerlaw} normalization. kT$_{plasma}$ and N$_{bremss}$ are plasma temperature and the {\tt bremss} normalization. E$_{gauss}$ and W$_{eq}$ are the Fe emission line energy and the equivalent width respectively. E$_{gabs}$ and $\sigma_{gabs}$ are the energy and width of the Gaussian absorption line in keV while $Depth_{gabs}$ is the line depth. F$_{0.3-3}$, F$_{3-6}$, F$_{6-10}$, F$_{10-20}$ and F$_{20-80}$ are fluxes in the energy range 0.3-3, 3-6, 6-10, 10-20 and 20-80 keV respectively.  }
\end{center}
\label{fitparm}
\end{table*}

\begin{figure}
\includegraphics[scale=0.30,angle=-90]{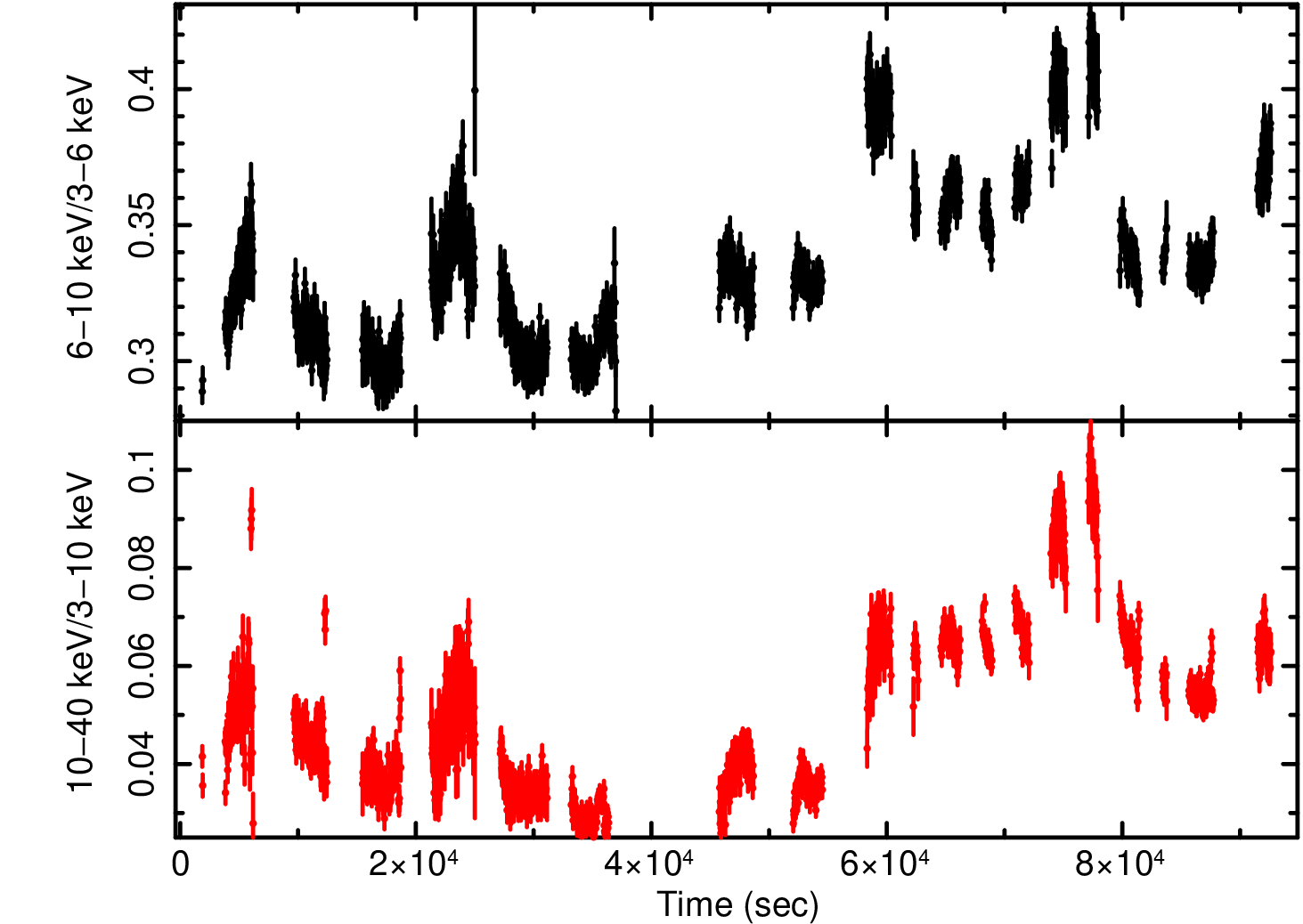}
\includegraphics[scale=0.30,angle=-90]{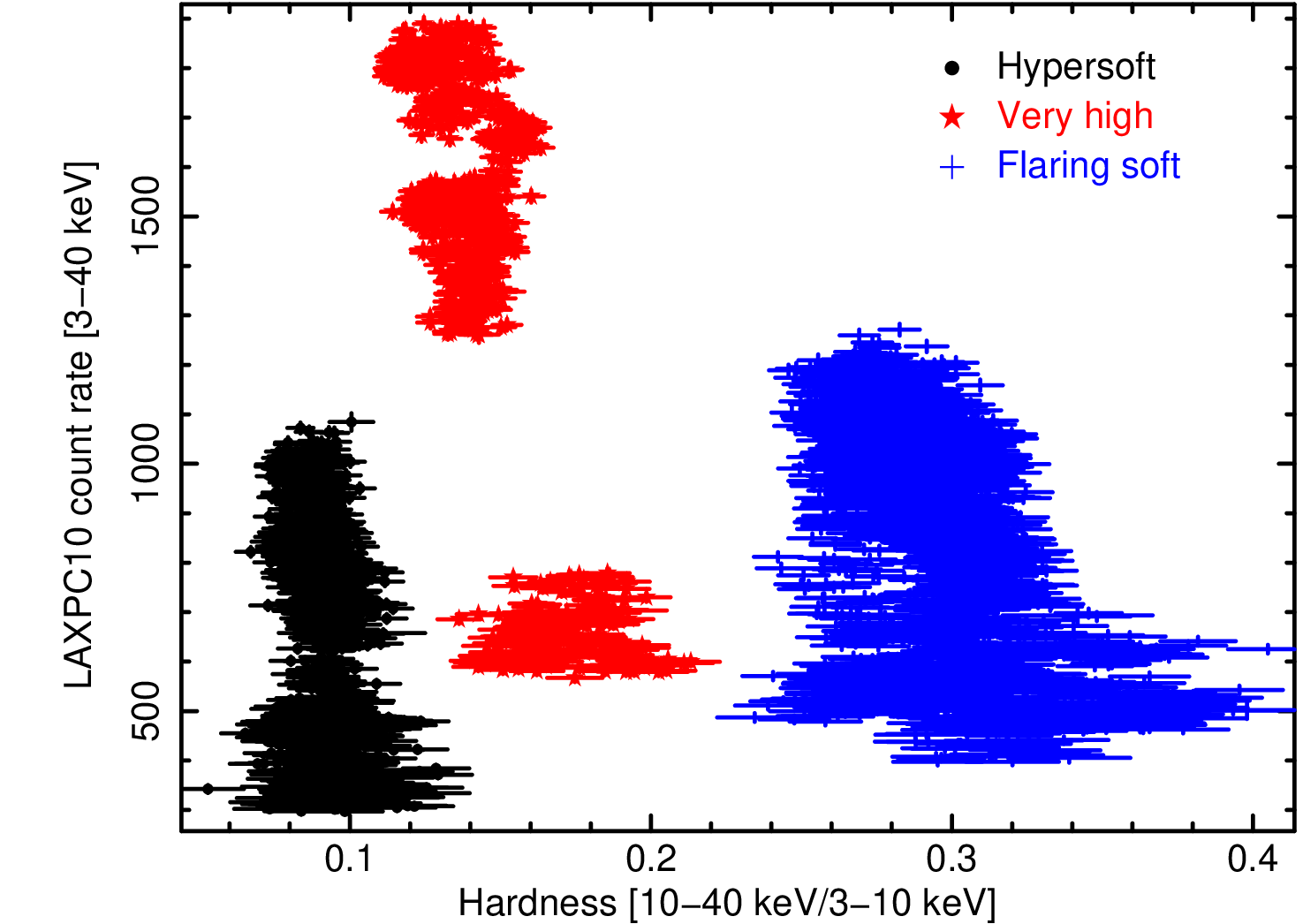}
\caption{Top left and bottom left panels show the soft color (ratio of count rate in 6-10 keV and 3-6 keV) and the hard color (ratio of count rate in 10-40 keV and 3-10 keV) as a function time. Right panel shows the hardness intensity diagram (HID) during the hypersoft (shown in black) and the very high (shown in red) state. Two separate structures in the HID of VHS at different count rate are due to the peak and dip phase count rate of the binary orbital motion in the presence of data gap. To compare with the HID from \citet{ko10}, we put the HID of the FSXR (shown in blue). The relative position of the FSXR and the hypersoft state are similar to that from \citet{ko10}, but a new state (very high state) is found between them with the \asat{} observation which coincide with the jetline proposed by \citet{ko10}.}
\label{hardness}
\end{figure}

\section{Observations and analysis}

\asat{} continuously observed Cyg X-3 between 01 April 2017 14:47:24 and 02 April 2017 16:43:39 covering 14 consecutive satellite orbits. Total observation duration is 93.3 ks. For the broadband spectroscopic purpose, we use simultaneous observations from the SXT and LAXPC instruments.

SXT is a focusing telescope with CCD camera that can perform X-ray imaging and spectroscopy in 0.3-8.0 keV energy range \citep{si14}. Level-1 imaging mode data is processed through a pipeline software to produce level-2 event files and then a good time interval (GTI) correcter and SXT event merger script\footnote{\label{note1}\url {http://www.tifr.res.in/~astrosat_sxt/page1_data_analysis.php}} is used to create merged event files from all the clean events with the corrected exposure time. We use {\textsc XSELECT V2.4d} in {\textsc HEASOFT 6.22.1} to extract lightcurve and spectra using source regions between 1 and 13 arcmin. An off-axis auxiliary response file (ARF) appropriate for the source location on the CCD is generated from the provided on-axis ARF using {\tt sxtmkarf} tool\footnoteref{note1}. Blank sky SXT observations are used to extract background spectrum. 

LAXPC consists of three identical X-ray proportional counters with the absolute time resolution of 10 $\mu$s in the energy range 3.0-80.0 keV \citep{ya16a, ya16b, mi16, an17, ag17}. LAXPC data are analyzed using the LAXPC software\footnote{\label{note2}\url {http://www.tifr.res.in/~astrosat_laxpc/LaxpcSoft.html}}. Details of the response and background spectra generation can be found in \citet{an17}. 

In the top left panel of Figure \ref{light}, we show the \swift{}/BAT daily lightcurve in 15-50 keV where the \asat{} observation span is marked by two vertical lines. The lightcurve shows that the \swift{}/BAT count rate sharply rises by a factor of $\sim$2 within the \asat{} observation period. The bottom panel shows the 11.2 GHz radio lightcurve obtained from RATAN-600 telescope \citep{tr17b}. The radio flux density was at a low, persistent level of $\sim$50-100 mJy until MJD 57845 and then it rises to $\sim$1 Jy within the \asat{} observation window. Therefore, the \asat{} observation during the onset of the giant Radio flare in simultaneous with the X-ray hardening is extremely valuable to monitor the disk/jet connections.

3-80 keV background-subtracted LAXPC lightcurve and 0.3-8 keV SXT lightcurve are shown in the top right panel of Figure \ref{light}. We note that the peak count rate of the binary modulation of $\sim$4.8 hours observed from both LAXPC and SXT lightcurves increases significantly ($\sim$50\%) after $\sim$60 ks. To check the soft and hard X-ray behavior of the source, we plot the 3-10 keV, 17-60 keV background-subtracted lightcurve and their ratio in the bottom left panel of Figure \ref{light}. With the 5\% background systematics, the source count rate in the 17-60 keV energy range is close to the non-detection level up to $\sim$60 ks and then increases significantly up to $\sim$150 counts/sec. Additionally, the hard-to-soft ratio becomes significant after $\sim$60 ks. This implies the absence of a hard component in the spectra during first 60 ks. This part of the observation belongs to the HPS. The hard component appears from 60 ks onwards during VHS. We check the flux-flux plot in 3-6 keV and 15-60 keV using LAXPC10 observations shown in the bottom right panel of Figure \ref{light}. During the HPS, 3-6 keV count rate strongly correlates with 15-60 keV count rate with the Spearman rank correlation coefficient (SRCC) of 0.82 (two-tailed p value is $<$ 0.0000004) while both lightcurve are not correlated during the VHS with the SRCC of 0.17 (two-tailed p value is $\sim$ 0.0006). The non-correlation indicates that the origin of hard and soft X-ray emission during the VHS may be independent of each other.  

\begin{figure*}
\begin{center}
\includegraphics[scale=0.33,angle=-90]{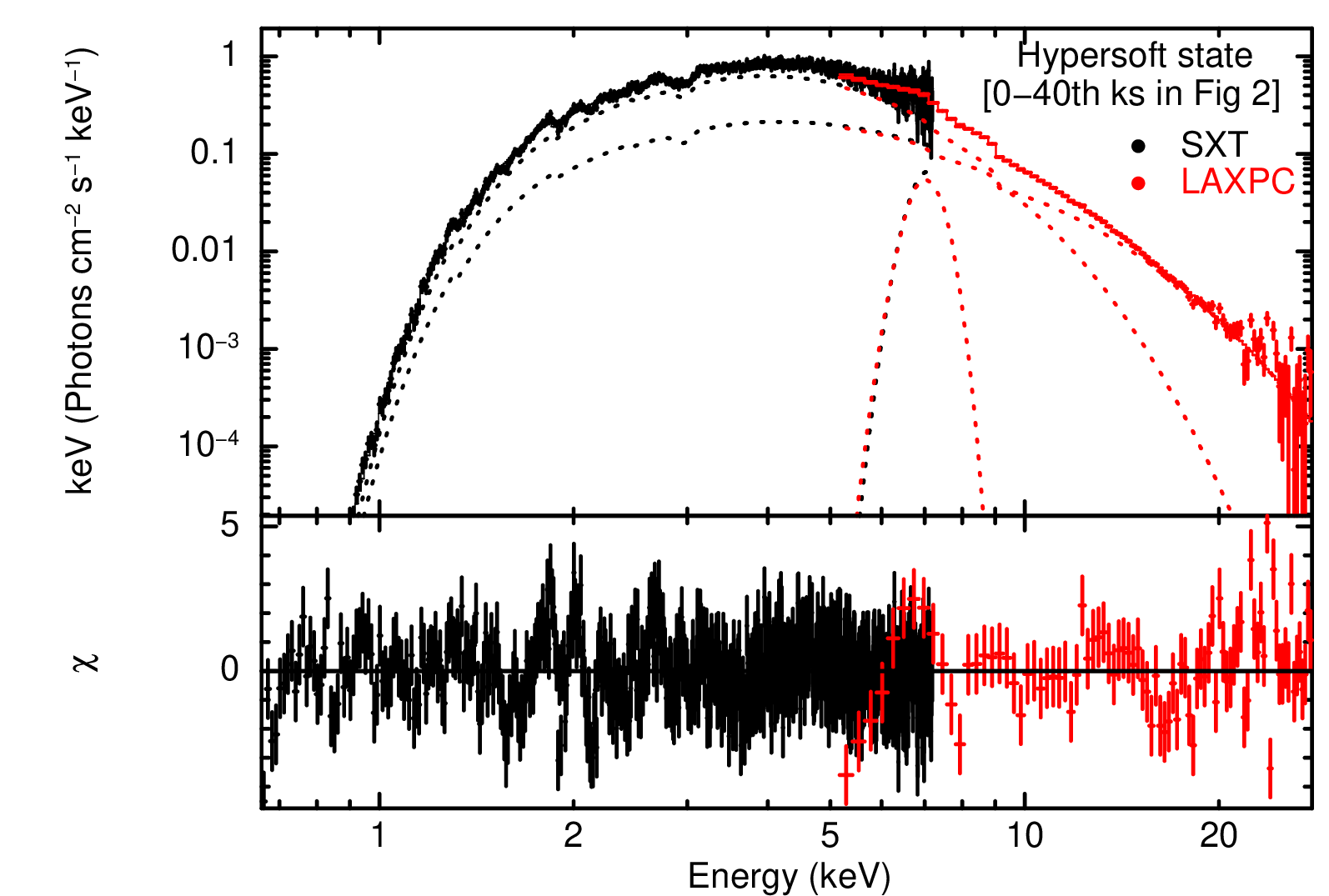}
\includegraphics[scale=0.33,angle=-90]{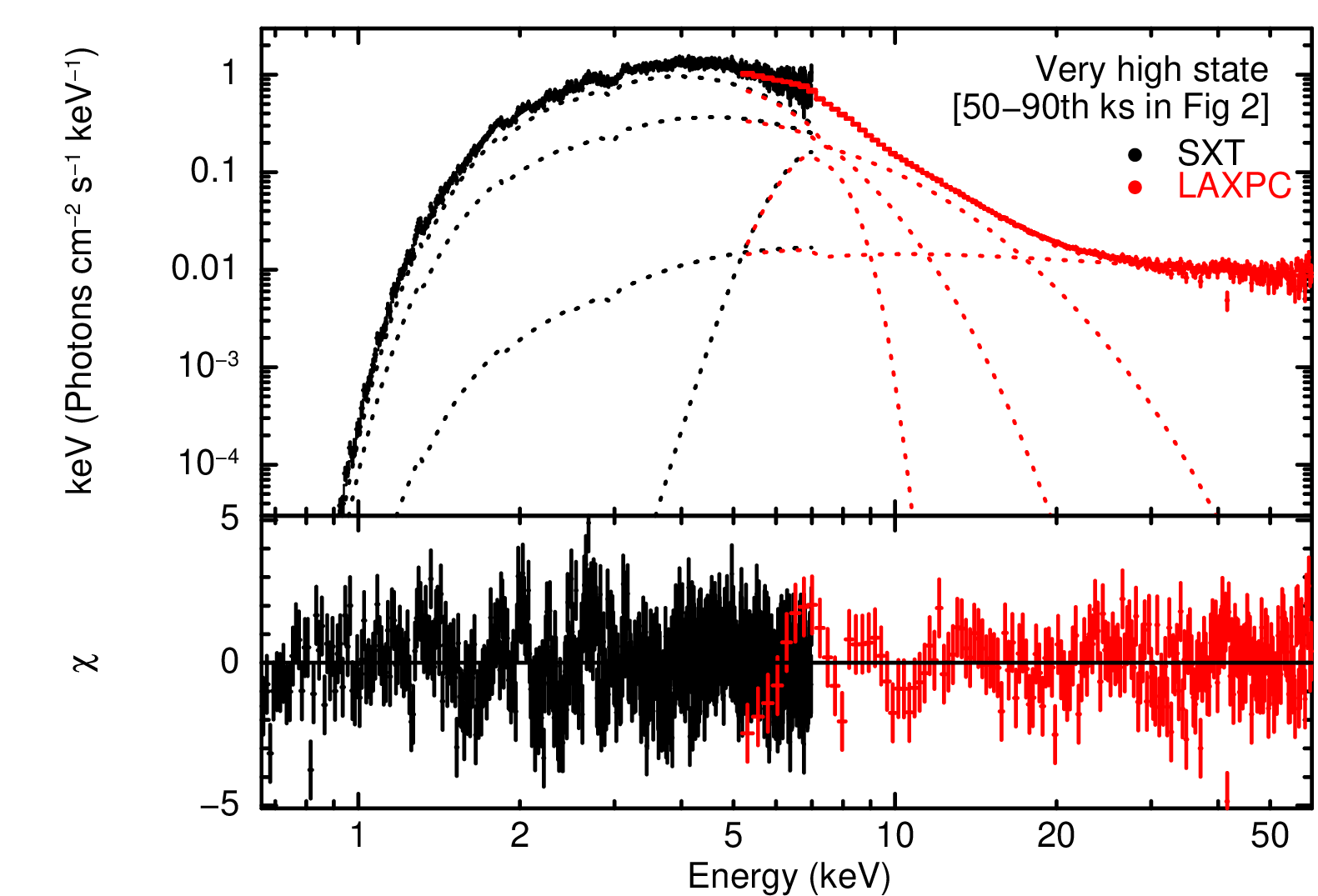}
\includegraphics[scale=0.33,angle=-90]{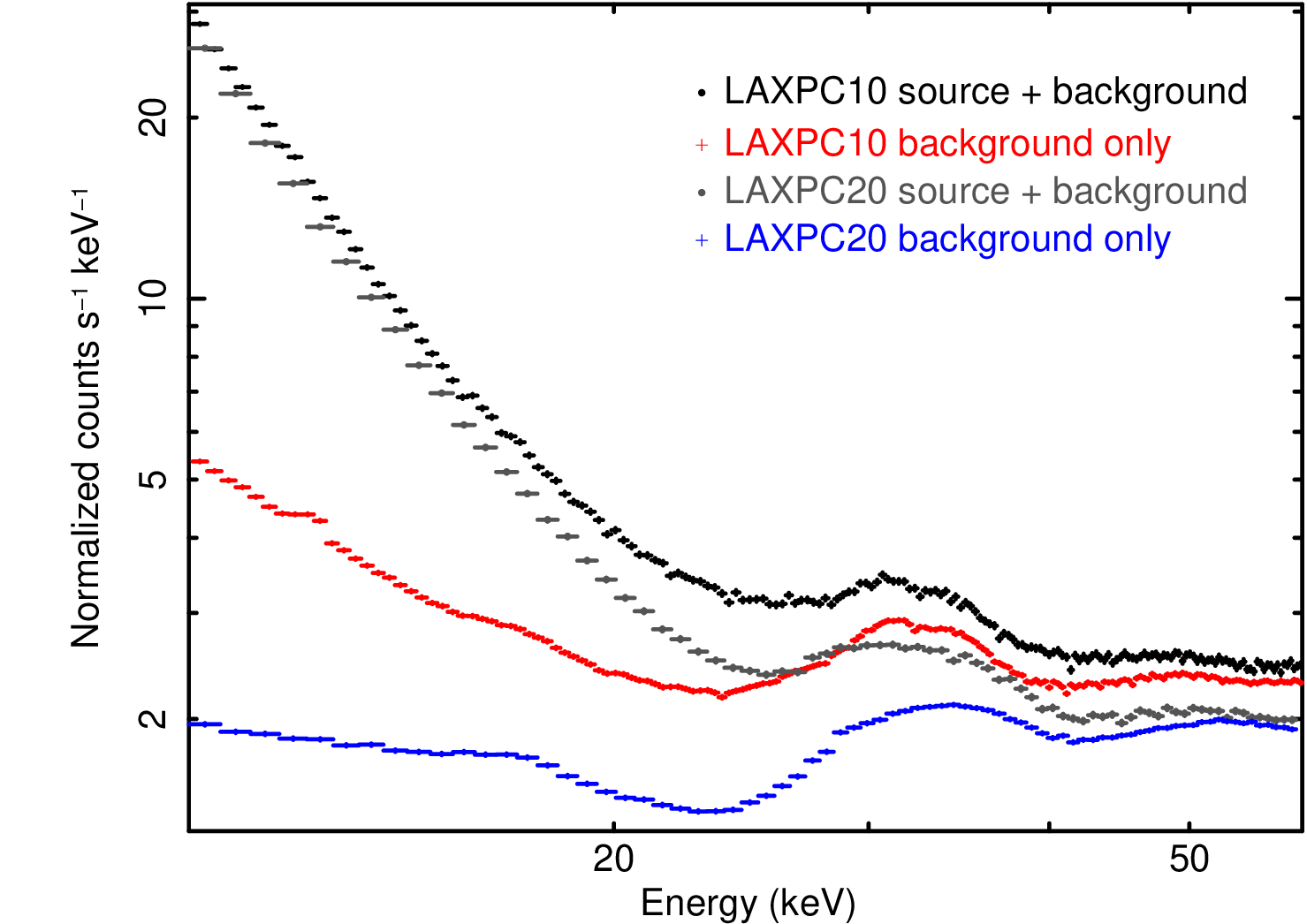}
\includegraphics[scale=0.33,angle=-90]{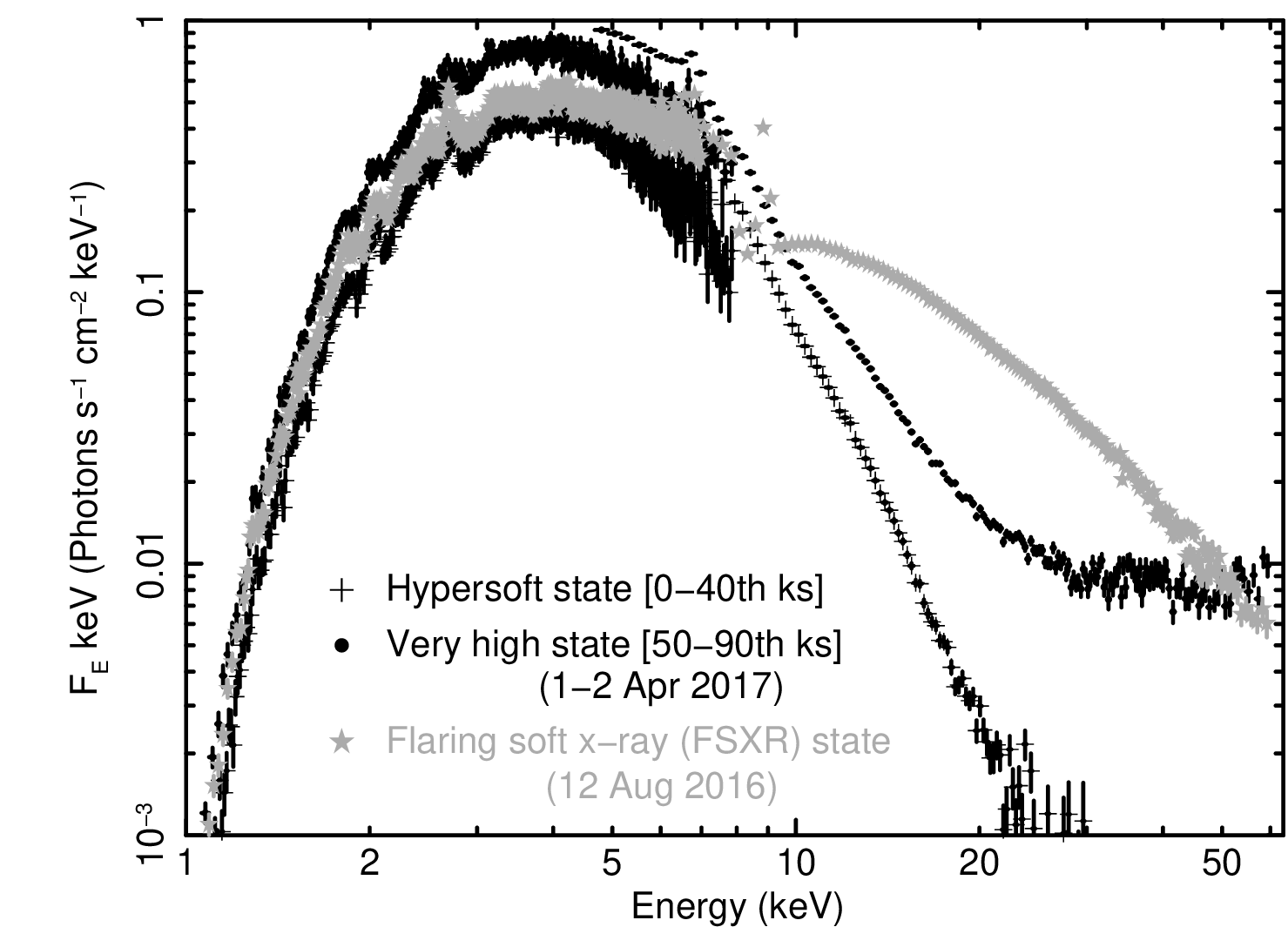}
\caption{Top left and top right panels show the best-fit joint spectra from SXT (black) and LAXPC (red) during the hypersoft and the very high states respectively along with the fitted model components and residuals. Bottom left panel shows the LAXPC10 and LAXPC20 source+background and background spectra simultaneously in 10-60 keV during the very high state. Due to the gain instability issue caused by a minor gas leakage, LAXPC30 spectra are excluded \citep{an17}. Bottom right panel shows the best-fit unfolded spectrum during the hypersoft (shown by pluses) and very high state (shown by circles). To compare these spectra with the flaring soft X-ray state (FSXR), we show the unfolded FSXR spectrum (shown in grey) which was observed on 12 August 2016 with \asat{}. }  
\label{spectra}
\end{center}
\end{figure*}

To examine the source variability, we plot the soft color (the ratio of count rate between 6-10 keV and 3-6 keV) and the hard color (the ratio of count rate in 10-40 keV and 3-10 keV) as a function of time in the left panel of figure \ref{hardness}. The modulations in the time-scale of the binary orbital period is seen in both soft and hard colors during the HPS and VHS. We confirm this by phase folding the data over the binary orbital period. The hardness intensity diagram (HID) plotted in the right panel of figure \ref{hardness} shows that the second half of the observation (shown in red) is significantly harder and brighter than the first half (shown in black) which motivates the naming as the `very high state'. For comparison, we also plot the HID of the FSXR (in blue) which was observed by \asat{} on 12 August 2016. 

\section{Spectral analysis and results}       

To confirm the HPS to VHS transition as indicated by the model-independent approach, we separately fit SXT+LAXPC joint spectra in the energy range 0.6-70.0 keV extracted during first 40 ks (HPS) and last 40 ks (VHS). We use the {\tt diskbb} model to fit the HPS emission \citep{ko10}. However, the disk is embedded in a plasma medium \citep{ma02,zd10} accreted from the companion, and therefore the bremsstrahlung emission from the collision between the inflowing and outflowing plasma \citep{ko13} has been taken care by the {\tt bremss} model. To account the interstellar medium absorption and the local, partial absorption by the accreted wind and hot medium, we use {\tt tbabs} and {\tt pcfabs} models respectively \citep{ko10}. Fitting with this continuum model, we found weak/no residual above 20 keV in the HPS but a strong residual in the VHS which resembles a powerlaw like emission. To account this hard X-ray residual, we use {\tt powerlaw} model in {\tt XSpec}. The Fe emission line and an absorption feature are modelled using a {\tt gauss} and a {\tt gabs} model respectively. Best-fit parameters along with 1$\sigma$ errors are provided in Table \ref{fitparm}. Best fit spectra during the HPS and the VHS along with their model components and residuals are shown in the top left and top right panels of Figure \ref{spectra}. In the bottom left panel, we plot source+background and background spectra during the VHS in 10-60 keV to show that the flat powerlaw like emission observed in the VHS is due to the source only. From Table \ref{fitparm}, we note that during the VHS, the hard X-ray excess has a photon powerlaw index of 1.43$^{+0.06}_{-0.03}$. Properties of such a flat powerlaw was not reported from any spectral states in Cyg X-3 and unlikely to be associated with the coronal emission. To make this point clear, we plot the energy spectra of the FSXR state (in gray) along with the best-fit HPS (shown by pluses) and VHS (shown by circles) in the bottom right panel of Figure \ref{spectra}. Comparing three spectra, we note that the FSXR spectrum has a very distinct and clear coronal emission component in the energy range 10-60 keV which is absent from the HPS and VHS. While the FSXR spectral fitting details will be discussed elsewhere, here it is used for comparison purpose. Since the time of giant radio flare formation coincides with the VHS, the most likely origin of the flat powerlaw component is the synchrotron emission from the jet base.

\subsection{Consistency check on the radio flux density from the X-ray spectra}

\begin{figure}
\begin{center}
\includegraphics[height=10 cm, width=6 cm, angle=-90]{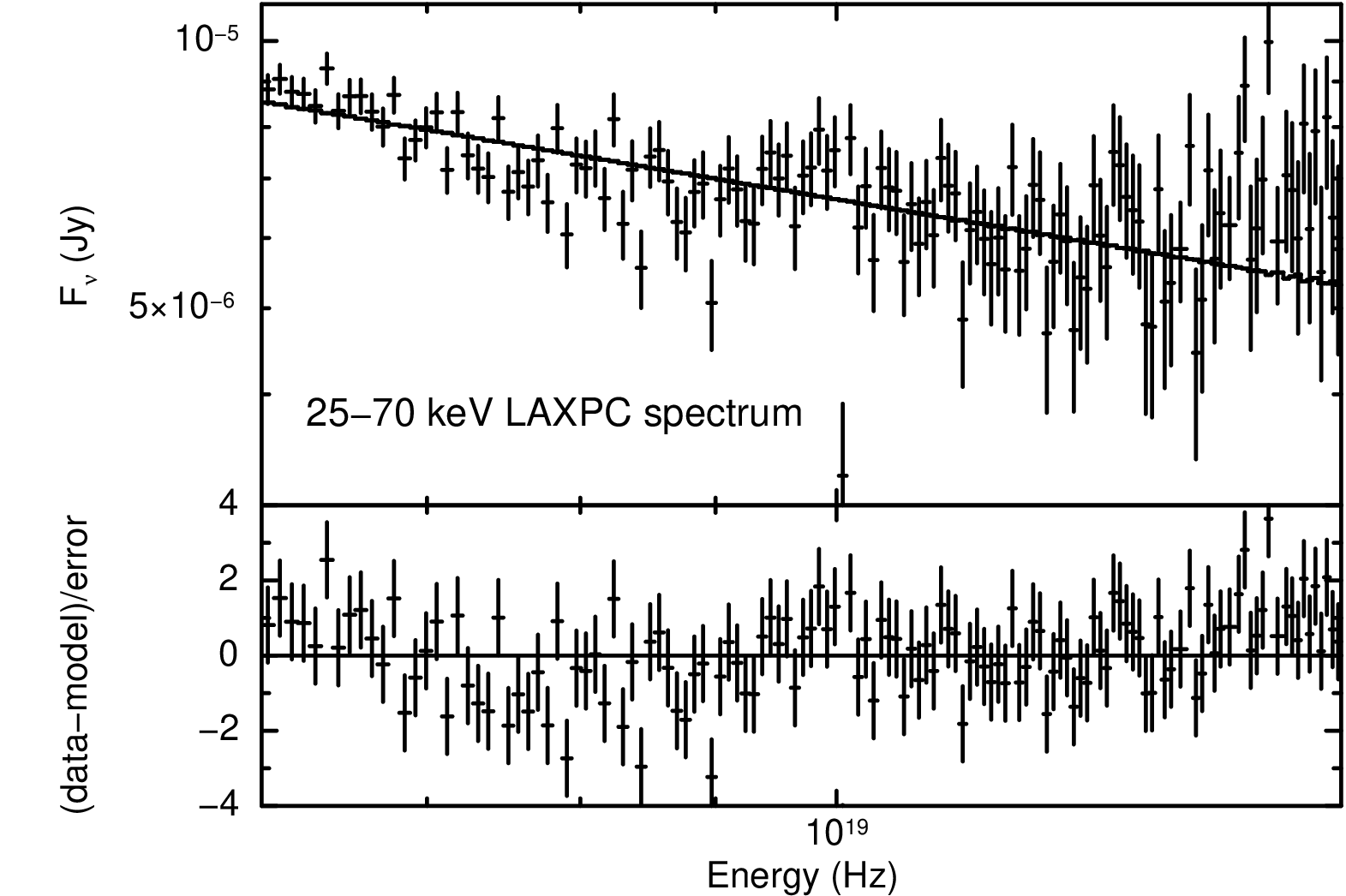}
\includegraphics[height=10 cm, width=6 cm, angle=-90]{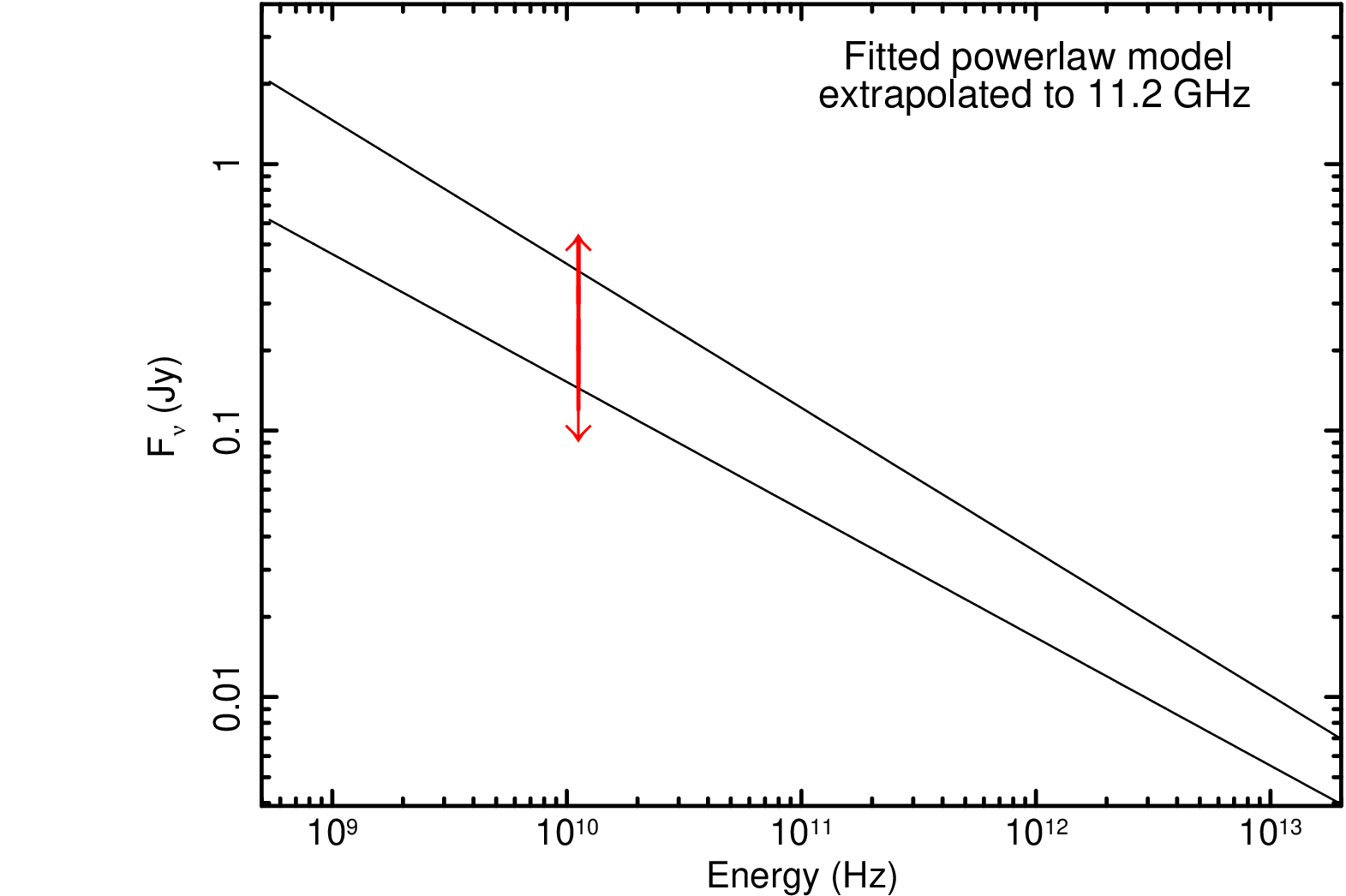}
\caption{The top and middle panels show 25-70 keV LAXPC spectra fitted with a powerlaw during the very high state and its residual respectively. The bottom panel shows the best-fit powerlaw model when extrapolated to radio frequencies. Two black lines are models correspond to the 2$\sigma$ upper and lower limits of powerlaw indices respectively. Vertical red line segment shows the range between two radio flux density measurements at 11.2 GHz during the X-ray observation of VHS.}
\end{center}
\label{radio}
\end{figure}

 If the synchrotron emission from the radio jet is the origin of the flat X-ray powerlaw, then an extrapolation of this powerlaw to the radio frequency should indicate the expected radio flux density. To check the consistency, we consider LAXPC spectrum of the VHS in 25-70 keV and fit it with a powerlaw. Such a choice of energy range ensures that no other model components contribute significantly in this band. The fitted spectrum and its residual are shown in the top and middle panels of Figure \ref{radio}. The best-fit powerlaw index and powerlaw normalization are 1.49$^{+0.04}_{-0.03}$ and 0.065$^{+0.006}_{-0.005}$ respectively. Both parameters are consistent with those from the powerlaw model used during the broadband spectral fitting of VHS (see Table \ref{fitparm}). When we extrapolate the fitted powerlaw model to the radio band, shown in the bottom panel of Figure \ref{radio}, we find that the radio flux density predicted by the X-ray powerlaw model in the range of $\sim$185-511 mJy at 11.2 GHz. The radio flux density observed by the RATAN-600 radio telescope at 11.2 GHz is 40 $\pm$ 2 mJy on MJD 57845.23 and 1650 $\pm$ 70 mJy on MJD 57846.21. The VHS spectrum analyzed here was observed between MJD 57845.34 and 57845.69. Therefore, our observation lie exactly between two radio observations and the X-ray spectral model fit predicted radio flux density is highly consistent with the trend shown in the top left panel of Figure \ref{light}. 

\section{Summary and conclusions}

In this work, we jointly analyze observations of Cyg X-3 using simultaneous SXT and LAXPC data on-board \asat{} on 01-02 April 2017 when the formation of a giant radio flare was reported using RATAN-600 radio telescope. 

During the first half of the \asat{} observation ($\sim$0-50 ks), the X-ray spectra is unusually soft, and the source is not detected significantly above 20 keV. This is consistent with the HPS which was previously observed few times from Cyg X-3 \citep{ko10}. The hardness ratio during the HPS is lower when compared to the FSXR state with no significant counts above 20 keV. 
During the second half of the observation ($\sim$60-93 ks), 20-60 keV count rate increases by a factor of $\sim$5-6 and the hardness ratio also increases by $\sim$50\%. This makes source spectral state different than that observed during the first half of the observation. Comparing with the FSXR, we show that the second half of \asat{} observation falls between the HPS and the FSXR and we termed here as the VHS. Interestingly if we qualitatively compare our HID with the Figure 3 from \citep{ko10}, we notice that the jet line which divide the HPS from FSXR in \citep{ko10} coincides with the position of the VHS as observed by the \asat{}. Therefore using \asat{} observations, we classify VHS as an additional state which falls between the HPS and FSXR state in the HID.   

An indication that the significant hard X-ray emission ($>$ 15 keV) during the VHS originates from a physical process different than the disk blackbody comes from a significantly low Spearman rank correlation coefficient (0.17) between 3-6 keV and 15-60 keV count rate while the same during the HPS is very high (0.82). The spectral modelling of the hard X-ray excess during the VHS indicates the powerlaw index of $\sim$1.49. The extrapolated powerlaw model predicts the radio flux density in the range of $\sim$185-511 mJy consistent with the trend observed from the RATAN-600 telescope at 11.2 GHz. Comparing with the contemporary radio flux we conclude that such a flat X-ray powerlaw emission is solely due to the synchrotron emission from the slowly moving jet base. The radio flare rise time is slower in Cyg X-3 \citep{tr17a} in comparison to the rise time seen in GRS 1915+105 \citep{ya06} which is consistent with slower speed of superluminal radio jets in Cyg X-3 \citep{ma00}. The detection of hard X-ray emission from radio jets in GRS 1915+105 may be difficult since accretion disk emission is stronger in GRS 1915+105 and superluminal radio jets in GRS 1915+105 are weaker at least by a factor over $\sim$20 \citep{ya06}.

From Table \ref{fitparm}, we note that during the HPS to the VHS transition, the inner disk temperature decreases from $\sim$1.39 keV to $\sim$1.11 keV while the inner disk radius (which is proportional to the square root of {\tt diskbb} normalization) increases by a factor of $\sim$2. This implies the inner disk recedes during the formation of a major plasma blob that emits radio flare. A significant increase in the contribution of the bremsstrahlung emission from $\sim$13\% (during HPS) to $\sim$36\% (during VHS) of the total flux in 0.1-100.0 keV along with appearance of the hard X-ray tail implies that the plasma blob, in the presence of a strong magnetic field, is capable of emitting synchrotron radiation in radio frequency and we detect the high energy part of such radiation as a flat powerlaw component in the X-ray spectra.

\section{Acknowledgement}  

We thank the referee for his/her constructive comments that improve the manuscript quality. We acknowledge the strong support from Indian Space Research Organization (ISRO) during instrument building, testing, software development and mission operation. We also acknowledge the support from the LAXPC Payload Operation Center (POC) and SXT POC at the TIFR, Mumbai.

\label{lastpage}

\end{document}